\documentclass[conference]{IEEEtran}
\IEEEoverridecommandlockouts
\usepackage{cite}
\usepackage{amsmath,amssymb,amsfonts}
\usepackage{algpseudocode}
\usepackage{tikz}
\usetikzlibrary{arrows.meta,positioning,fit,calc}
\usepackage{graphicx}
\usepackage{textcomp}
\usepackage{xcolor}
\usepackage{siunitx}
\usepackage{multicol}
\usepackage{multirow}
\usepackage{lipsum}
\usepackage{url}
\usepackage{subfig}
\usepackage{epstopdf}
\usepackage{float}
\usepackage{dblfloatfix} 

\usepackage[linesnumbered,lined,boxed,commentsnumbered,ruled,vlined]{algorithm2e}
\graphicspath{ {Images/} }

\def\BibTeX{{\rm B\kern-.05em{\sc i\kern-.025em b}\kern-.08em
    T\kern-.1667em\lower.7ex\hbox{E}\kern-.125emX}}

\begin{document}

\title{GhostLock: Learning-Assisted Blind GPS Signal Nullification Without Victim-Side Information
}

\author{\IEEEauthorblockN{
Khalil Ahmad\IEEEauthorrefmark{1}$^\ddagger$,
Omar Ansari\IEEEauthorrefmark{1}$^\ddagger$,
Mohsin Kamal\IEEEauthorrefmark{1},
Asif Iqbal\IEEEauthorrefmark{2},
and Muhammad Naveed Aman\IEEEauthorrefmark{1}
}
\IEEEauthorblockA{
\IEEEauthorrefmark{1}\textit{School of Computing, University of Nebraska--Lincoln}, Lincoln, NE 68508, USA\\
\IEEEauthorrefmark{2}\textit{National University of Singapore}, Singapore\\
\{kahmad2, hansari2\}@huskers.unl.edu\\
\{mkamal3, naveed.aman\}@unl.edu\\
aiqbal@nus.edu.sg
}
}

\maketitle

\begin{abstract}
The Global Positioning System (GPS) provides critical positioning, navigation, and timing (PNT) services for numerous safety- and mission-critical applications, making it an attractive target for physical-layer attacks. While GPS spoofing and jamming have been extensively studied, GPS Signal Nullification (GSN)-a physical-layer attack that suppresses authentic satellite signals through destructive interference prior to acquisition-remains largely unexplored. Existing wireless signal nullification techniques target conventional communication systems and typically require victim-side information, such as channel estimates or synchronization parameters, making them impractical for GPS. Furthermore, many GPS spoofing defenses assume authentic satellite signals remain available as trusted references, an assumption invalidated by GSN. To address this gap, we present GhostLock, a learning-assisted blind GSN framework requiring only an approximate target location estimate. GhostLock integrates a Blind Adaptive Filter (BAF) with a deep learning-based synchronization module to generate and align an anti-signal using only the attacker's locally received GPS signal, eliminating the need for victim-side observations. We implement GhostLock on a real-time software-defined radio (SDR) platform using GPS-SDR-SIM and GNU Radio. Experimental results demonstrate successful suppression of GPS signal acquisition, establishing the feasibility of blind GSN and motivating new detection and mitigation techniques for this emerging GPS threat model.

\end{abstract}

\begin{IEEEkeywords}
GPS Signal Nullification, Blind Adaptive Filtering, Deep Learning, Physical-Layer Security, CNN
\end{IEEEkeywords}

\section{Introduction}

\IEEEPARstart{T}{he} Global Positioning System (GPS) provides positioning, navigation, and timing (PNT) services that underpin numerous critical infrastructures, including transportation, power grids, telecommunications, financial networks, autonomous vehicles, and military systems \cite{ogobor2025global,salim2024cybersecurity}. Because civilian GPS signals arrive at Earth with extremely low received power after propagating over approximately 20,000 km, they are highly vulnerable to intentional interference such as jamming, spoofing, and signal nullification \cite{tryb2025gnss}. Successful attacks can degrade navigation accuracy, disrupt timing synchronization, and impair safety-critical operations across interconnected cyber-physical systems \cite{prezelj2025global,solanki2025survey}.

\begin{figure}
    \centering
    \includegraphics[width=\linewidth]{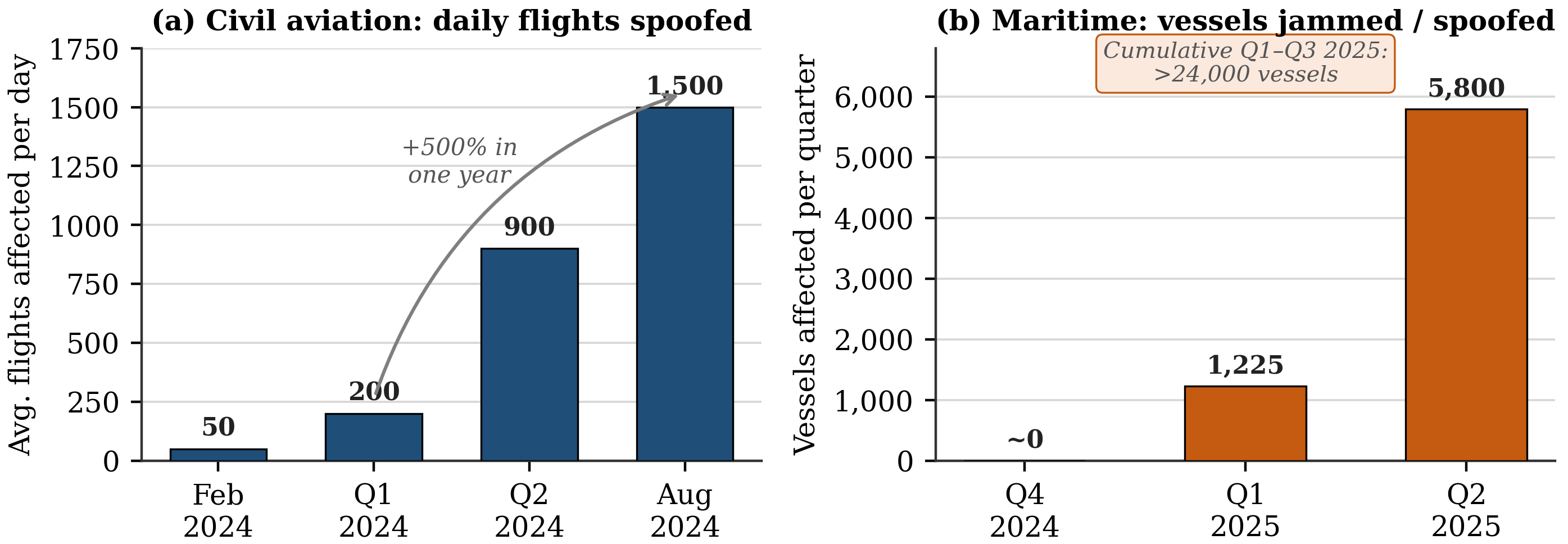}
    \caption{Reported GPS interference incidents across critical-infrastructure sectors.}
    \label{fig:intro1}
\end{figure}

Recent years have witnessed a sharp increase in GPS interference incidents. Maritime authorities have reported thousands of disrupted vessels, while aviation agencies have documented a dramatic rise in flights affected by GPS spoofing and jamming \cite{westbrook2023radiofrequency,felux2024impacts,okochi2025gps}. Figure \ref{fig:intro1} summarizes this growing trend across major infrastructure sectors. In particular, the average number of aircraft affected by spoofing increased from only a few dozen per day in early 2024 to approximately 1,500 flights per day by August 2024 \cite{opsgroup2024gps}, while more than 24,000 maritime vessels experienced GPS interference during the first three quarters of 2025 \cite{Windward2025MaritimeStressTest}. These incidents demonstrate that GPS interference is evolving into a widespread threat to critical infrastructure.

\begin{figure}
    \centering
    \includegraphics[width=\linewidth]{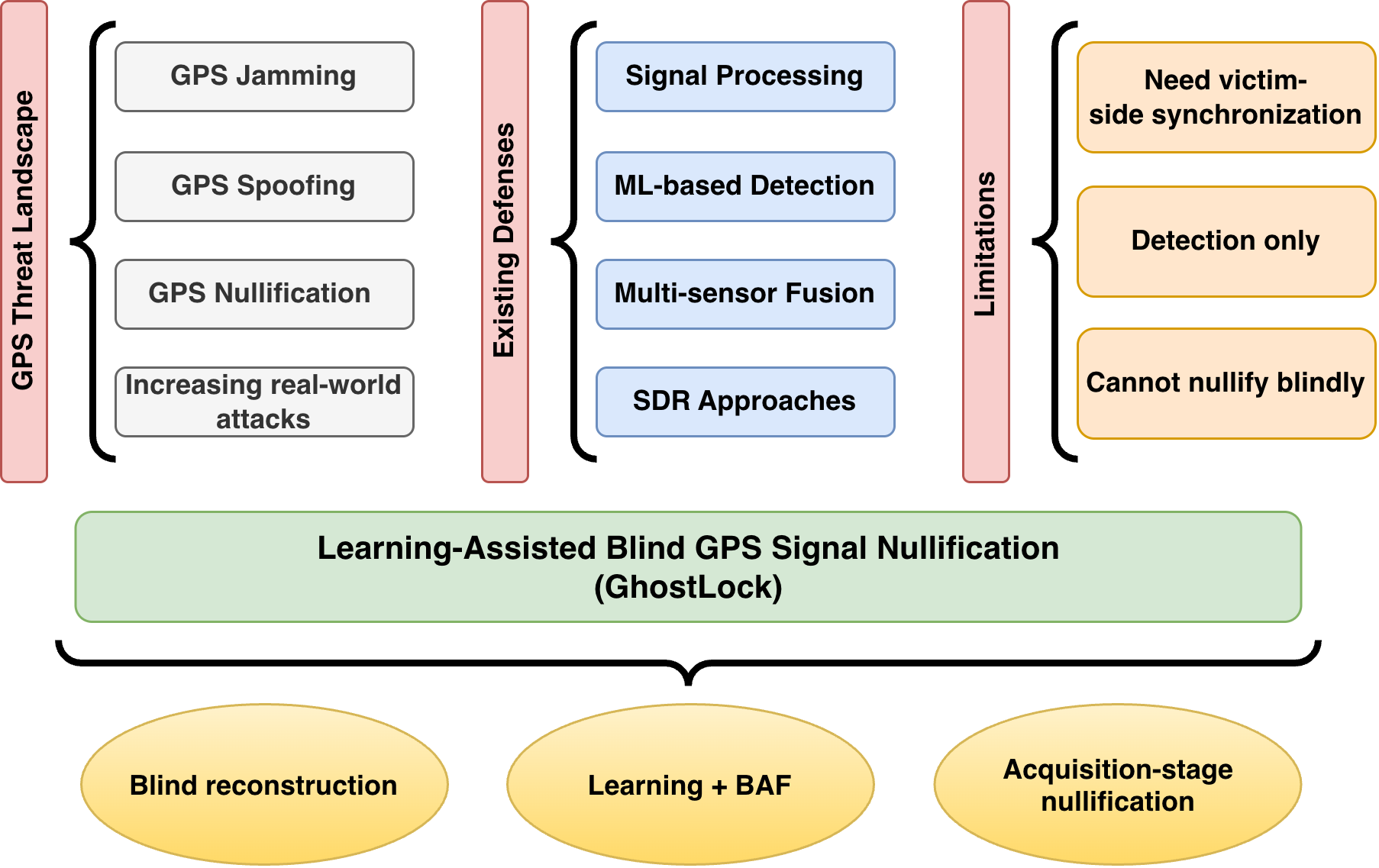}
    \caption{Evolution of GPS attack threats, existing defense approaches, and the research gap addressed by the proposed GhostLock framework.}
    \label{fig:motivation}
\end{figure}

Although GPS spoofing and jamming have received significant attention, GPS signal nullification remains comparatively underexplored. Figure \ref{fig:motivation} summarizes the current research landscape by contrasting existing attack types, representative defense techniques, and their limitations. While prior signal-processing and machine-learning approaches have substantially improved spoofing detection, they largely rely on victim-side observations or focus solely on attack detection. Consequently, blind GPS signal nullification under minimal attacker knowledge remains an open research problem, motivating the development of the proposed \textit{GhostLock} framework. 
As illustrated in Figure \ref{fig:motivation}, existing solutions primarily address spoofing detection rather than the synthesis of coherent cancellation signals capable of preventing GPS signal acquisition. Moreover, most approaches assume some degree of synchronization or receiver-side information, assumptions that are often impractical in realistic attack scenarios.

Existing research has primarily followed two complementary directions. The first direction focuses on signal-processing-based techniques that analyze physical-layer characteristics of GNSS signals, including carrier phase, correlation functions, and synchronization parameters, to detect or mitigate spoofing attacks. For example, Friedt et al. \cite{friedt2021real} proposed a software-defined radio (SDR)-based framework that exploits multi-antenna carrier-phase analysis for real-time spoofing detection and mitigation, while Gao et al. \cite{gao2023new} designed an asynchronous spoofing algorithm by carefully controlling the code rate, carrier frequency, and signal power to capture receiver tracking loops. More recently, the auhtors in \cite{liu2025efficient} introduced a complex correlator-based spoofing detection framework that combines multiple signal-quality metrics derived from the in-phase and quadrature components of GPS signals. In parallel, machine-learning-based approaches have emerged as a promising alternative by learning complex temporal and statistical relationships from GNSS observations and auxiliary sensor data. Shabbir et al. \cite{shabbir2023securing} and Dasgupta et al. \cite{dasgupta2022sensor} proposed deep learning and multi-sensor fusion frameworks for GPS spoofing detection, whereas Alqahtani et al. \cite{alqahtani2026deep} introduced the Transformer-Gated GPS Spoofing Detection (TG-GSD) model that directly analyzes raw GPS signal features. Although these approaches have substantially improved spoofing detection and receiver resilience, they primarily focus on identifying attacks rather than generating coherent cancellation signals capable of preventing GPS signal acquisition.

\begin{figure}
    \centering
    \includegraphics[width=\linewidth]{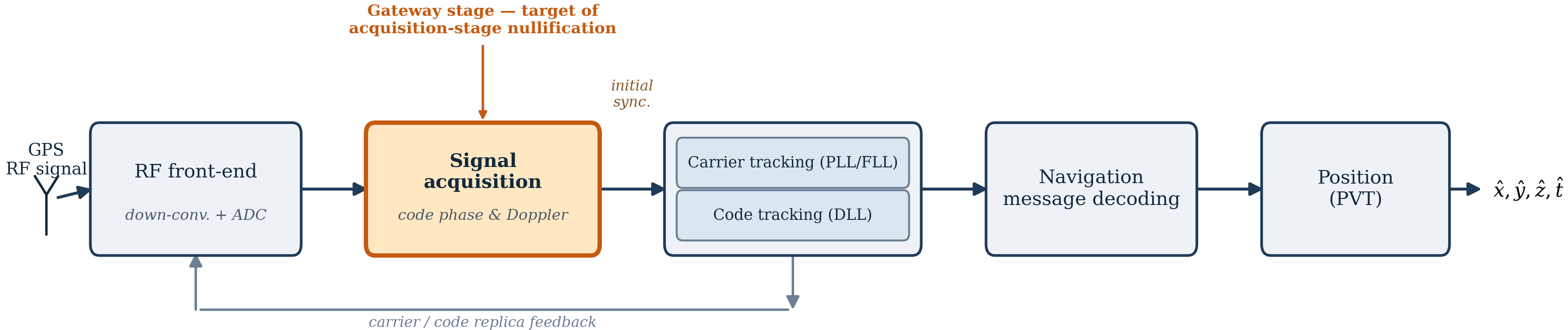}
    \caption{Signal-processing chain of a typical GPS receiver, from RF front-end through acquisition, carrier and code tracking, navigation-message decoding, and position (PVT) estimation.}
    \label{fig:intro2}
\end{figure}

A typical GPS receiver consists of RF front-end processing, signal acquisition, carrier and code tracking, navigation-message decoding, and position estimation, as illustrated in Figure \ref{fig:intro2}. Among these stages, signal acquisition is the critical gateway because it establishes the initial code-phase and Doppler synchronization required for all subsequent receiver operations. Preventing successful acquisition effectively disables tracking and navigation regardless of downstream receiver algorithms \cite{psiaki2016gnss,elmiger2025gnss}. This observation motivates targeting the acquisition stage for GPS signal nullification.

In this paper, we present \textit{GhostLock}, a learning-assisted blind GPS signal nullification framework that requires only a coarse estimate of the victim's geographical location. Unlike existing techniques that depend on victim-side information or synchronization feedback, GhostLock combines a machine-learning-based synchronization estimator with blind adaptive filtering to reconstruct coherent GPS anti-signals capable of suppressing authentic satellite signals before successful acquisition. By demonstrating GPS signal nullification under minimal attacker knowledge, GhostLock exposes a previously unexplored threat model and motivates the development of stronger GPS protection mechanisms.

The main contributions of this paper are summarized as follows:

\begin{itemize}

\item We propose \textit{GhostLock}, the first learning-assisted blind GPS signal nullification framework that operates without victim-side observations, requiring only a coarse estimate of the victim's location.

\item We develop a machine-learning-based synchronization module that predicts the delay and phase required for coherent GPS signal cancellation, together with a blind adaptive filtering framework that reconstructs the anti-signal for acquisition-stage nullification.

\item We implement GhostLock on an SDR-based experimental platform and extensively evaluate its performance against existing GPS nullification approaches, demonstrating effective signal cancellation under realistic attacker constraints.

\end{itemize}

\section{System Model and Assumptions}
\label{sec:system_model}
\begin{figure}[t!]
    \centering
    \includegraphics[width=\linewidth]{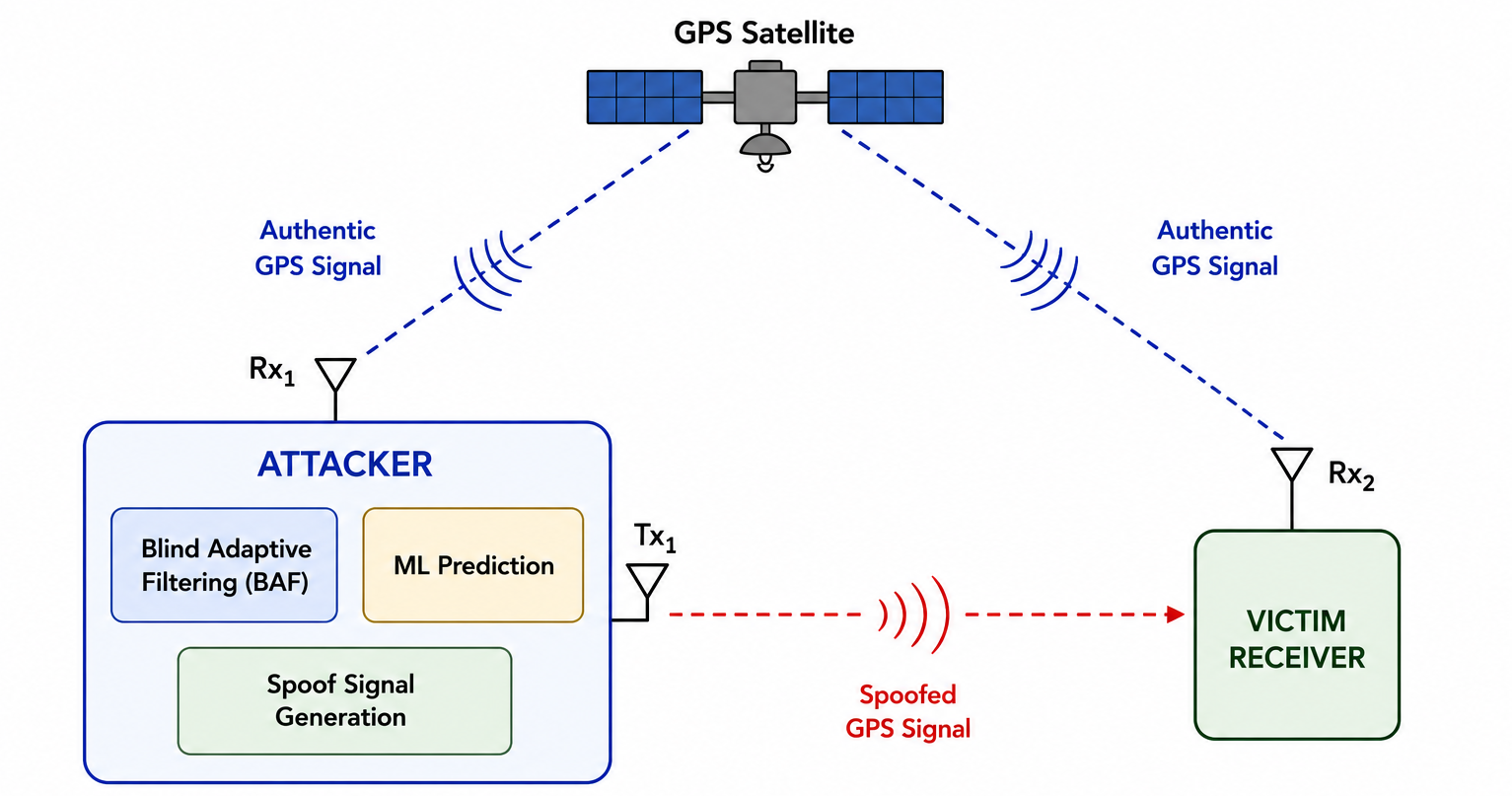}
    \caption{System block diagram.}
    \label{fig:system_model}
\end{figure}

Let's consider the learning-assisted GSN framework illustrated in Figure \ref{fig:system_model}. The system comprises: (i) a set of visible GPS satellites, (ii) a passive victim GPS receiver, (iii) an attacker GPS observation receiver, (iv) a BAF module, (v) a learning-based synchronization module, and (vi) an SDR-based RF transmitter. The attack operates in the analog RF domain before the victim performs signal acquisition. The attacker first receives the authentic GPS signals at its own antenna and uses a BAF to reconstruct the locally observed GPS waveform with inverted phase. A ML model then predicts the carrier-phase corrections required to transform the attacker-side waveform into a cancellation waveform appropriate for the estimated victim location. The corrected anti-signal is then transmitted towards the victim.

After propagating from the attacker to the victim, the RF cancellation signal combines over the air with the authentic satellite signals at the victim antenna. Thus, the waveform entering the victim RF front end is the analog residual resulting from destructive interference of RF signals. The victim subsequently downconverts and digitizes this residual waveform. If the authentic GPS components have been sufficiently attenuated, the subsequent receiver acquisition process cannot detect the corresponding satellite correlation peaks. The complex-envelope and discrete-time representations introduced below are used to derive the attacker-side signal processing operations and to implement them on an SDR. They do not imply that cancellation occurs digitally inside the victim receiver. The physical nullification event occurs only after RF transmission, at the victim antenna and RF front end.

\subsection{Authentic GPS RF Signal Model}
\label{subsec:gps_rf_signal_model}
Let's consider $K$ visible GPS satellites. The authentic GPS RF signal arriving at location $q\in\{a,v\}$, where $a$ and $v$ denote the attacker and victim, respectively, can be modeled as:

\begin{equation}
\begin{aligned}
s_q^{\mathrm{RF}}(t)
= &
\sum_{k=1}^{K}
A_{q,k}(t)
C_k\!\left(t-\tau_{q,k}(t)\right)
D_k\!\left(t-\tau_{q,k}(t)\right)
\\ &
\times \cos\!\left(
2\pi\left(f_c+f_{D,q,k}(t)\right)t
+
\phi_{q,k}(t)
\right),   
\end{aligned}
\label{eq:gps_rf_signal}
\end{equation}
where $A_{q,k}(t)$ is the received amplitude of satellite $k$, $C_k(t)$ is its publicly known spreading code, $D_k(t)$ is its navigation-data sequence, $\tau_{q,k}(t)$ is the effective signal delay, $f_c$ is the GPS carrier frequency, $f_{D,q,k}(t)$ is the received Doppler shift, and $\phi_{q,k}(t)$ is the carrier phase at location $q$. For compact analysis, the equivalent complex envelope is

\begin{equation}
\begin{aligned}
z_q(t)
= &
\sum_{k=1}^{K}
A_{q,k}(t)
C_k\!\left(t-\tau_{q,k}(t)\right)
D_k\!\left(t-\tau_{q,k}(t)\right)
\\ &
\times e^{j\left(
2\pi f_{D,q,k}(t)t+\phi_{q,k}(t)
\right)}.  
\end{aligned}
\label{eq:gps_complex_envelope}
\end{equation}
The physical RF waveform is recovered from the complex envelope as:

\begin{equation}
s_q^{\mathrm{RF}}(t)
=
\Re\left\{
z_q(t)e^{j2\pi f_ct}
\right\}.
\label{eq:complex_envelope_to_rf}
\end{equation}
Equation~\eqref{eq:gps_complex_envelope} is therefore an analytical
representation of the RF waveform rather than a model of cancellation inside the victim's digital receiver. The attacker samples its locally received complex envelope at sampling period $T_s$, producing

\begin{equation}
x_a[n]
=
z_a(nT_s)+\nu_a[n],
\label{eq:attacker_sampled_signal}
\end{equation}
where $\nu_a[n]$ denotes attacker receiver noise, quantization error, and RF front-end distortion.

\subsection{Spatial Difference Between Attacker and Victim Signals}
\label{subsec:spatial_signal_difference}

Although the attacker and victim may observe the same visible satellites and navigation-data bits, the received GPS waveforms are not identical because the antennas occupy different spatial locations. For satellite $k$, we define:

\begin{align}
\Delta\tau_k(t)
&=
\tau_{v,k}(t)-\tau_{a,k}(t),
\label{eq:relative_satellite_delay}
\\
\Delta f_{D,k}(t)
&=
f_{D,v,k}(t)-f_{D,a,k}(t),
\label{eq:relative_satellite_doppler}
\\
\Delta\phi_k(t)
&=
\phi_{v,k}(t)-\phi_{a,k}(t),
\label{eq:relative_satellite_phase}
\\
g_k(t)
&=
\frac{A_{v,k}(t)}{A_{a,k}(t)}.
\label{eq:relative_satellite_gain}
\end{align}
An idealized transformation between the attacker- and victim-side satellite components may then be written as:

\begin{equation}
z_{v,k}(t)
=
g_k(t)
z_{a,k}\!\left(t-\Delta\tau_k(t)\right)
e^{j\left(
2\pi\Delta f_{D,k}(t)t+\Delta\phi_k(t)
\right)}.
\label{eq:attacker_victim_satellite_mapping}
\end{equation}
GhostLock does not calculate the transformations in \eqref{eq:attacker_victim_satellite_mapping} using an explicit analytical distance model. Instead, the relevant phase corrections are learned from paired measurements collected at different attacker-victim separations and deployment conditions.

\subsection{Attacker-to-Victim RF Transmission}
\label{subsec:attacker_victim_transmission}

Let $x_c^{\mathrm{RF}}(t)$ denote the RF cancellation waveform transmitted by the attacker. Before reaching the victim, it undergoes propagation delay, attenuation, carrier-phase rotation, and RF hardware distortion. We represent these combined effects using an effective attacker-to-victim RF operator $\mathcal{H}_{av}\{\cdot\}$:

\begin{equation}
s_{c,v}^{\mathrm{RF}}(t)
=
\mathcal{H}_{av}
\left\{
x_c^{\mathrm{RF}}(t)
\right\}.
\label{eq:effective_rf_operator}
\end{equation}
For interpretability, a dominant-path representation can be expressed as:

\begin{equation}
s_{c,v}^{\mathrm{RF}}(t)
=
\alpha_{av}(t)
x_c^{\mathrm{RF}}
\!\left(t-\tau_{av}(t)\right)
\Big|_{\mathrm{phase\ rotated}}
+
\zeta_{av}^{\mathrm{RF}}(t),
\label{eq:effective_dominant_path}
\end{equation}
where $\alpha_{av}(t)$ is the effective amplitude scaling, $\tau_{av}(t)$ is the RF propagation delay, and $\zeta_{av}^{\mathrm{RF}}(t)$ represents multipath and unmodeled transmission effects. Neither $\alpha_{av}(t)$ nor $\tau_{av}(t)$ is assumed to be known through direct victim feedback. Their combined effect on the required transmit waveform is incorporated into the synchronization corrections learned from prior measurements.

\subsection{Analog RF Residual at the Victim}
\label{subsec:analog_rf_residual}
The waveform physically presented to the victim antenna and RF front end is:

\begin{equation}
r_v^{\mathrm{RF}}(t)
=
s_v^{\mathrm{RF}}(t)
+
s_{c,v}^{\mathrm{RF}}(t)
+
n_v^{\mathrm{RF}}(t),
\label{eq:victim_rf_residual_additive}
\end{equation}
where $n_v^{\mathrm{RF}}(t)$ denotes thermal noise and residual power due to cancellation mismatches. The attacker designs the cancellation waveform such that:

\begin{equation}
s_{c,v}^{\mathrm{RF}}(t)
\approx
-s_v^{\mathrm{RF}}(t).
\label{eq:rf_anti_signal_condition}
\end{equation}
Under ideal amplitude, delay, frequency, data-bit, and carrier-phase alignment,

\begin{equation}
r_v^{\mathrm{RF}}(t)
\approx
n_v^{\mathrm{RF}}(t).
\label{eq:ideal_rf_residual}
\end{equation}
In the remainder of the paper, we use the subtractive form for analytical clarity and insert the physical $\pi$ inversion only once during RF anti-signal generation.

\subsection{System Assumptions}
\label{subsec:system_assumptions}
We adopt the following system assumptions:

\subsubsection{Passive and Stationary Victim}
The victim is assumed to be a conventional GPS receiver that does not transmit pilot signals, channel state information. Furthermore, the victim is assumed to remain stationary during a single observation, such that the attacker-victim channel is quasi-static over the adaptation window. This assumption is well justified for many GPS-dependent applications, including fixed infrastructure, timing receivers, and parked vehicles, and is commonly adopted as a baseline in experimental GPS spoofing research before considering dynamic receivers \cite{ranganathan2016spree}.

\subsubsection{Approximate Victim Location}
The attacker is assumed to possess an approximate estimate of the victim's location. Exact localization or sub-wavelength position estimation is not required. This assumption is reasonable in many practical scenarios where the target's coarse location is known through prior observation \cite{kang2018adaptive, kamal2021gps}.

\subsubsection{No Online Victim-Side Feedback}
The attacker has no online access to the victim's raw I/Q samples, residual signals or correlator outputs. Consequently, all online adaptation relies solely on attacker-side GPS measurements, approximate victim location, and synchronization model. This assumption distinguishes the proposed attack from closed-loop spoofing approaches that require victim-side observations or feedback \cite{tippenhauer2011requirements}.





\section{Threat Model}
\label{sec:threat_model}

\subsection{Adversarial Objective}
\label{subsec:adversarial_objective}

We consider a sophisticated adversary seeking to suppress authentic GPS signals through analog RF destructive interference. The immediate objective is GPS RF signal nullification. Nullification may subsequently enable a spoofing stage, but the present work focuses on suppressing the authentic signals before the victim can reliably acquire them.

The attacker seeks to synthesize an RF waveform whose propagated version approximates the negative of the authentic composite GPS waveform at the victim antenna:

\begin{equation}
\mathcal{H}_{av}
\left\{
x_c^{\mathrm{RF}}(t)
\right\}
\approx
-s_v^{\mathrm{RF}}(t).
\label{eq:attacker_rf_objective}
\end{equation}
Equivalently, in complex-envelope notation, the attacker seeks to minimize:

\begin{equation}
J_v
=
\mathbb{E}
\left[
\left|
z_v(t)-\widetilde{z}_v(t)
\right|^2
\right],
\label{eq:true_victim_residual_objective}
\end{equation}
where $\widetilde{z}_v(t)$ denotes the positive-phase estimate of the cancellation waveform arriving at the victim. Its RF implementation is transmitted with a $\pi$ phase inversion.





\subsection{Attacker Capabilities}
\label{subsec:attacker_capabilities}

The attacker has following capabilities:

\begin{enumerate}
    \item \textit{Attacker-side GPS reception:}
    The attacker can receive authentic GPS signals at its own location using an SDR-based RF front end.

    \item \textit{Local parameter estimation:}
    The attacker can estimate locally observable quantities such as satellite identity, code delay, Doppler shift, amplitude, and carrier phase.

    \item \textit{Offline paired-data collection:}
    During system development, the attacker can collect paired attacker- and target-location measurements at multiple distances and propagation conditions. Victim-side measurements are used only during offline training and evaluation, not during the online attack.

    \item \textit{Learning-based synchronization:}
    A trained ML model can predict the carrier phase corrections needed for locally reconstructed waveform to arrive at the victim in destructive alignment.

    \item \textit{Adaptive attacker-side waveform reconstruction:}
    The attacker can use an adaptive filter to reconstruct the authentic GPS waveform from locally available observations and reference signals.

    \item \textit{Controlled RF synthesis and transmission:}
    The attacker can generate an RF anti-signal with controllable amplitude, phase, and frequency using an SDR transmitter.
\end{enumerate}

\subsection{Attacker Limitations}
\label{subsec:attacker_limitations}

During online operation, the attacker does not know the exact instantaneous waveform at the victim. In particular, the attacker
cannot observe:

\begin{enumerate}
    \item the victim-side analog residual;
    \item the victim's raw or digitized I/Q samples;
    \item the victim's instantaneous RF channel;
    \item the victim's acquisition or correlator outputs;
    \item any acknowledgment or feedback indicating cancellation success.
\end{enumerate}
The standard BAF based update is infeasible as victim's signal is unavailable to the attacker. GhostLock instead adapts the BAF using a locally measurable attacker-side reconstruction error. The learned synchronization module subsequently maps the locally reconstructed waveform to the estimated victim location.

\section{Proposed Technique: GhostLock}
\label{sec:proposed_technique}

GhostLock is a learning-assisted blind GPS signal-nullification framework that operates without online victim-side information. It contains the following stages:

\begin{enumerate}
    \item attacker-side GPS signal observation
    \item local BAF-based waveform reconstruction;
    \item ML prediction of victim-specific phase correction;
    \item generation of anti-signal; and
    \item analog destructive superposition at the victim antenna.
\end{enumerate}

The division of functions is important. The BAF estimates and reconstructs the authentic waveform available at the attacker. It does not adapt against an unavailable victim residual. The learning module predicts how the locally reconstructed waveform must be shifted in phase before RF transmission so that, after propagation, it opposes the authentic GPS waveform at the victim.

\subsection{Attacker-Side GPS Observation and Reference Construction}
\label{subsec:attacker_observation}

The attacker records complex samples $x_a[n]$ as defined in \eqref{eq:attacker_sampled_signal}. Using its locally available GPS observations, the attacker constructs a reference vector:

\begin{equation}
\mathbf{u}[n]
=
\begin{bmatrix}
u_1[n] &
u_2[n] &
\cdots &
u_M[n]
\end{bmatrix}^{T},
\label{eq:baf_reference_vector}
\end{equation}
where each $u_m[n]$ represents a delayed reference sample. For example, a tapped-delay-line reference can be written as:

\begin{equation}
\mathbf{u}[n]
=
\begin{bmatrix}
x_a[n] &
x_a[n-1] &
\cdots &
x_a[n-M+1]
\end{bmatrix}^{T}.
\label{eq:tapped_delay_reference}
\end{equation}
Alternatively, when satellite components are separately reconstructed,

\begin{equation}
\mathbf{u}[n]
=
\begin{bmatrix}
u_1[n;\widehat{\boldsymbol{\vartheta}}_{a,1}] &
\cdots &
u_K[n;\widehat{\boldsymbol{\vartheta}}_{a,K}]
\end{bmatrix}^{T},
\label{eq:satellite_reference_vector}
\end{equation}
where

\begin{equation}
\widehat{\boldsymbol{\vartheta}}_{a,k}
=
\begin{bmatrix}
\widehat{A}_{a,k} &
\widehat{\tau}_{a,k} &
\widehat{f}_{D,a,k} &
\widehat{\phi}_{a,k}
\end{bmatrix}^{T}.
\label{eq:local_parameter_estimates}
\end{equation}

\subsection{Blind Adaptive Reconstruction at the Attacker}
\label{subsec:baf_reconstruction}

The BAF output is given by

\begin{equation}
\widehat{z}_a[n]
=
\mathbf{w}^{H}[n]\mathbf{u}[n],
\label{eq:baf_attacker_output}
\end{equation}
where $\mathbf{w}[n]\in\mathbb{C}^{M}$ is the complex adaptive-weight vector, $\mathbf{u}[n]$ is the reference vector. Unlike a formulation that requires the unknown victim-side residual, GhostLock uses the locally observable attacker-side reconstruction error:

\begin{equation}
e_a[n]
=
x_a[n]-\widehat{z}_a[n]
=
x_a[n]-\mathbf{w}^{H}[n]\mathbf{u}[n].
\label{eq:attacker_local_error}
\end{equation}
The BAF coefficients are selected to minimize the attacker-side mean-square reconstruction error. 

\begin{equation}
J_a(\mathbf{w})
=
\mathbb{E}
\left[
\left|
x_a[n]-\mathbf{w}^{H}\mathbf{u}[n]
\right|^2
\right].
\label{eq:attacker_reconstruction_objective}
\end{equation}
The corresponding mean-square reconstruction objective is:

\begin{equation}
\mathbf{w}_{\mathrm{opt}} = \arg\min_{\mathbf{w}} J_a(\mathbf{w}).
\label{eq:w_opt}
\end{equation}
Expanding \eqref{eq:attacker_reconstruction_objective} gives:

\begin{align}
J_a(\mathbf{w})
&=
P_a - \mathbf{w}^{H}\mathbf{p}_{ux} -
\mathbf{p}_{ux}^{H}\mathbf{w}
+
\mathbf{w}^{H}\mathbf{R}_{uu}\mathbf{w},
\label{eq:attacker_mse_expansion}
\end{align}
where

\begin{align}
P_a
&=
\mathbb{E}
\left[ |x_a[n]|^2
\right],
\label{eq:attacker_received_power}
\\
\mathbf{R}_{uu}
&=
\mathbb{E}
\left[
\mathbf{u}[n]\mathbf{u}^{H}[n]
\right],
\label{eq:baf_reference_covariance}
\\
\mathbf{p}_{ux}
&=
\mathbb{E}
\left[
\mathbf{u}[n]x_a^{*}[n]
\right].
\label{eq:baf_cross_correlation}
\end{align}
Using the complex gradient,

\begin{equation}
\frac{\partial J_a}
{\partial\mathbf{w}^{*}}
=
-\mathbf{p}_{ux}
+
\mathbf{R}_{uu}\mathbf{w}.
\label{eq:baf_complex_gradient}
\end{equation}
Setting the gradient to zero yields the Wiener solution:

\begin{equation}
\mathbf{w}_{\mathrm{opt}}
=
\mathbf{R}_{uu}^{-1}\mathbf{p}_{ux},
\label{eq:wiener_solution}
\end{equation}
The corresponding minimum attacker-side reconstruction error is:

\begin{equation}
J_a^{\min}
=
P_a
-
\mathbf{p}_{ux}^{H}
\mathbf{R}_{uu}^{-1}
\mathbf{p}_{ux}.
\label{eq:baf_wiener_bound}
\end{equation}
Equation~\eqref{eq:baf_wiener_bound} gives the best linear reconstruction of the attacker-observed signal achievable using the selected reference space. It does not, by itself, represent the minimum residual power at the victim because victim-side synchronization and RF transmission errors remain.

\begin{figure*}[t!]
\centering
\resizebox{\textwidth}{!}{%
\begin{tikzpicture}[
    node distance=5mm and 8mm,
    inputblock/.style={
        draw=blue!70!black,
        fill=blue!10,
        thick,
        rounded corners,
        align=center,
        minimum height=12mm,
        minimum width=25mm,
        font=\footnotesize
    },
    preprocessblock/.style={
        draw=cyan!70!black,
        fill=cyan!10,
        thick,
        rounded corners,
        align=center,
        minimum height=9mm,
        minimum width=21mm,
        font=\footnotesize
    },
    cnnblock/.style={
        draw=purple!70!black,
        fill=purple!10,
        thick,
        rounded corners,
        align=center,
        minimum height=9mm,
        minimum width=21mm,
        font=\footnotesize
    },
    smallblock/.style={
        draw=purple!70!black,
        fill=purple!6,
        thick,
        rounded corners,
        align=center,
        minimum height=8mm,
        minimum width=18mm,
        font=\scriptsize
    },
    processblock/.style={
        draw=orange!80!black,
        fill=orange!12,
        thick,
        rounded corners,
        align=center,
        minimum height=9mm,
        minimum width=21mm,
        font=\footnotesize
    },
    arrow/.style={
        -{Latex[length=2mm]},
        thick,
        draw=gray!75!black
    },
    branch/.style={
        -{Latex[length=2mm]},
        thick,
        draw=cyan!70!black
    }
]

\node[inputblock] (iq) {
I/Q window\\
$\mathbf{X}\in\mathbb{R}^{2\times N_w}$
};

\node[preprocessblock,right=of iq] (norm) {
Channel-wise\\
normalization
};

\node[cnnblock,right=of norm] (s1) {
Stage-1 CNN\\
full-delay regression
};

\node[smallblock,right=of s1] (round) {
Round\\
$\widehat m=\operatorname{round}(\widehat d_1)$
};

\node[cnnblock,right=of round] (s2) {
Stage-2 CNN\\
fractional refinement
};

\node[smallblock,right=of s2] (combine) {
Combine\\
$\widehat d=\widehat m+\widehat\alpha$
};

\node[processblock,right=of combine] (align) {
Fractional-delay\\
alignment
};

\node[processblock,right=of align] (phase) {
Delay-to-phase\\
conversion
};

\draw[arrow] (iq) -- (norm);
\draw[arrow] (norm) -- (s1);

\draw[arrow] (s1) --
node[above,font=\scriptsize,text=purple!70!black]
{$\widehat d_1$}
(round);

\draw[arrow] (round) --
node[above,font=\scriptsize,text=purple!70!black]
{$\widehat m$}
(s2);

\draw[arrow] (s2) --
node[above,font=\scriptsize,text=purple!70!black]
{$\widehat\alpha$}
(combine);

\draw[arrow] (combine) --
node[above,font=\scriptsize,text=purple!70!black]
{$\widehat d$}
(align);

\draw[arrow] (align) -- (phase);

\draw[branch]
(norm.south)
|- ++(0,-12mm)
-| (s2.south);

\node[
    font=\scriptsize,
    fill=white,
    text=cyan!70!black,
    inner sep=1pt,
    anchor=west
]
at ($(norm.south)+(2mm,-10mm)$)
{normalized I/Q};

\node[
    draw=purple!70!black,
    fill=purple!4,
    fill opacity=0.35,
    text opacity=1,
    dashed,
    thick,
    rounded corners,
    fit=(s1)(round)(s2)(combine),
    inner sep=3mm,
    label={
        [font=\footnotesize\bfseries,
         text=purple!70!black]
        above:Two-stage learned delay estimator
    }
] {};

\end{tikzpicture}%
}
\caption{Implemented two-stage learning pipeline. The input consists of the I/Q observation window and the attacker--victim separation distance $d$. Stage~1 estimates the full propagation delay and provides its rounded integer component to Stage~2. Stage~2 predicts the remaining fractional delay. The final delay estimate is then converted analytically into the carrier phase required for RF cancellation.}
\label{fig:two_stage_ml_pipeline}
\end{figure*}
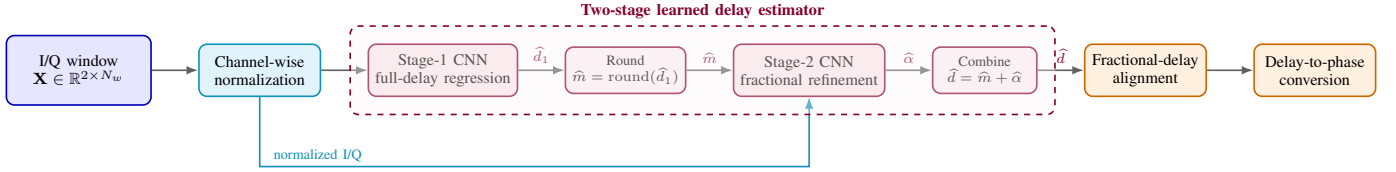

\subsection{Two-Stage Learning-Based Delay Prediction}
\label{subsec:two_stage_delay_prediction}

The BAF output provides the attacker-side waveform for cancellation-signal generation. To temporally align the reconstructed and authentic waveforms at the target, GhostLock uses a two-stage 1-D CNN to estimate the effective relative delay from complex baseband I/Q samples. Fig.~\ref{fig:two_stage_ml_pipeline} illustrates the proposed two-stage delay estimator. Stage~1 predicts the integer-delay component from the input I/Q signal, while Stage~2 uses the same signal and the Stage~1 output to estimate the fractional component. Both estimates are then combined to obtain the final sub-sample delay.

Unlike approaches that rely on hand-crafted spatial or satellite-specific features, the proposed predictor operates directly on a fixed-length window of complex baseband samples. For the \(i\)th training example, the network input is represented as

\begin{equation}
\mathbf{X}^{(i)}
=
\begin{bmatrix}
\Re\!\left\{x_a^{(i)}[0]\right\} &
\cdots &
\Re\!\left\{x_a^{(i)}[N_w-1]\right\}
\\
\Im\!\left\{x_a^{(i)}[0]\right\} &
\cdots &
\Im\!\left\{x_a^{(i)}[N_w-1]\right\}
\end{bmatrix}
\in \mathbb{R}^{2 \times N_w},
\label{eq:ml_iq_input}
\end{equation}

where \(x_a^{(i)}[n]\) denotes the \(n\)th complex attacker-side baseband sample in training example \(i\). The first and second rows of \(\mathbf{X}^{(i)}\) contain the in-phase and quadrature components, respectively, while \(N_w\) denotes the observation-window length.

Each input matrix \(\mathbf{X}^{(i)}\) is paired with a supervised delay label \(d^{(i)}\), expressed in samples. This label characterizes the effective relative temporal offset between the attacker-side waveform and the corresponding waveform measured at the calibration receiver.

\subsection{Delay-Label Decomposition}

To support both integer- and sub-sample temporal alignment, the target delay is decomposed as

\begin{equation}
d^{(i)}
=
m^{(i)}+\alpha^{(i)},
\qquad
m^{(i)}\in\mathbb{Z},
\quad
\alpha^{(i)}\in[-0.5,0.5],
\label{eq:delay_integer_fraction_decomposition}
\end{equation}

where \(m^{(i)}\) denotes the nearest integer-sample delay and \(\alpha^{(i)}\) denotes the residual fractional-sample offset. Equivalently, the two components may be defined as

\begin{equation}
m^{(i)}=\operatorname{round}\!\left(d^{(i)}\right),
\qquad
\alpha^{(i)}=d^{(i)}-m^{(i)}.
\label{eq:delay_components}
\end{equation}

The restriction \(\alpha^{(i)}\in[-0.5,0.5]\) follows from the nearest-integer definition of \(m^{(i)}\). This decomposition converts the original delay-estimation problem into two complementary tasks. The first stage determines the coarse integer-sample alignment, while the second stage estimates the remaining fractional offset. The final delay prediction is consequently reconstructed as

\begin{equation}
\widehat{d}^{(i)}
=
\widehat{m}^{(i)}+\widehat{\alpha}^{(i)},
\label{eq:reconstructed_delay}
\end{equation}

where \(\widehat{m}^{(i)}\) and \(\widehat{\alpha}^{(i)}\) are the integer- and fractional-delay estimates produced by the two network stages, respectively. This hierarchical formulation reduces the range of the fine-regression task and enables sub-sample temporal synchronization.

\subsubsection{Stage 1: Full-Delay Regression}

Following the decomposition in \eqref{eq:delay_integer_fraction_decomposition}, the first stage provides the coarse delay estimate required to determine the integer-sample alignment. Specifically, the Stage-1 network, denoted by \(f_{\Theta_1}(\cdot)\), maps the normalized I/Q observation window to a scalar full-delay estimate:

\begin{equation}
\widehat{d}_{1}^{(i)}
=
f_{\Theta_1}
\left(
\widetilde{\mathbf{X}}^{(i)}
\right),
\label{eq:stage1_delay_prediction}
\end{equation}

where \(\widetilde{\mathbf{X}}^{(i)}\) is obtained by independently standardizing the in-phase and quadrature channels using statistics computed exclusively from the training set. The resulting estimate \(\widehat{d}_{1}^{(i)}\) represents the predicted relative delay in samples prior to integer quantization.

The Stage-1 feature extractor begins with a one-dimensional convolution of kernel size \(15\) and stride \(2\), followed by group normalization and a SiLU activation. The resulting features are processed by eight residual convolutional blocks with kernel size \(7\) and cyclic dilation factors \(1\), \(2\), \(4\), and \(8\). Stride-\(2\) convolutional layers are inserted after the second, fourth, and sixth residual blocks to progressively reduce the temporal resolution while increasing the effective receptive field. Adaptive average pooling then converts the final feature map into a fixed-dimensional representation, which is passed through a multilayer regression head to generate \(\widehat{d}_{1}^{(i)}\).

During training, the delay targets are standardized using the training-set statistics, and the network parameters \(\Theta_1\) are optimized using the smooth-\(L_1\) objective

\begin{equation}
\mathcal{L}_{1}
=
\frac{1}{N_{\mathrm{tr}}}
\sum_{i=1}^{N_{\mathrm{tr}}}
\operatorname{SmoothL1}
\left(
\widehat{d}_{1}^{(i)},
d^{(i)}
\right),
\label{eq:stage1_smooth_l1_loss}
\end{equation}

where both the predictions and labels are evaluated in the standardized target domain during optimization. After transforming the Stage-1 output back to the original sample domain, the integer-delay estimate supplied to Stage 2 is obtained as

\begin{equation}
\widehat{m}^{(i)}
=
\operatorname{round}
\left(
\widehat{d}_{1}^{(i)}
\right).
\label{eq:stage1_integer_delay}
\end{equation}

Thus, Stage 1 resolves the coarse integer-sample alignment, while Stage 2 estimates the residual fractional component \(\widehat{\alpha}^{(i)}\). The two outputs are subsequently combined according to \eqref{eq:reconstructed_delay} to obtain the final sub-sample delay prediction.

\subsubsection{Stage 2: Fractional-Delay Refinement}

The second-stage network, denoted by $f_{\Theta_2}(\cdot)$, predicts the fractional correction relative to the rounded Stage-1 estimate. The training target is
\begin{equation}
\alpha^{(i)}
=
d^{(i)}-\widehat m^{(i)}.
\label{eq:fractional_delay_target}
\end{equation}
Training examples for which $\alpha^{(i)}\notin[-0.5,0.5]$ indicate an incorrect Stage-1 integer estimate and are considered invalid for fractional refinement.

The Stage-2 architecture contains two parallel feature paths. The first path is a residual downsampling trunk. It begins with a convolution of kernel size $15$ and stride $4$, followed by four residual downsampling blocks and an attention-pooling layer. The second path is a fine-resolution branch containing six dilated convolutions with dilation factors
\begin{equation}
\mathcal{D}_{2}
=
\{1,2,4,8,16,32\}.
\label{eq:stage2_dilation_set}
\end{equation}
The two feature vectors are concatenated and passed through a fully-connected regression head. To constrain the fractional output to its valid range, the final network output is
\begin{equation}
\widehat\alpha^{(i)}
=
\frac{1}{2}
\tanh
\left(
g_{\Theta_2}
\left(
\widetilde{\mathbf{X}}^{(i)}
\right)
\right),
\label{eq:fractional_delay_prediction}
\end{equation}
where $g_{\Theta_2}(\cdot)$ denotes the unconstrained output of the Stage-2 regression head. The final two-stage delay estimate is
\begin{equation}
\widehat d^{(i)}
=
\widehat m^{(i)}
+
\widehat\alpha^{(i)}.
\label{eq:final_two_stage_delay}
\end{equation}

\subsubsection{Cancellation-Aware Training Objective}

Stage~2 is trained with a composite loss that combines supervised delay regression with a physical cancellation term. The total objective is
\begin{equation}
\mathcal{L}_{2}
=
\lambda_{\alpha}\mathcal{L}_{\alpha}
+
\lambda_{d}\mathcal{L}_{d}
+
\lambda_{c}(e)\mathcal{L}_{c},
\label{eq:stage2_total_loss}
\end{equation}
where the first two terms are supervised smooth-$L_1$ losses on the fractional and total delay:
\begin{align}
\mathcal{L}_{\alpha}
&=
\frac{1}{N_{\mathrm{tr}}}
\sum_{i=1}^{N_{\mathrm{tr}}}
\operatorname{SmoothL1}
\left(
\widehat\alpha^{(i)},
\alpha^{(i)}
\right),\\
\mathcal{L}_{d}
&=
\frac{1}{N_{\mathrm{tr}}}
\sum_{i=1}^{N_{\mathrm{tr}}}
\operatorname{SmoothL1}
\left(
\widehat d^{(i)},
d^{(i)}
\right).
\end{align}
The cancellation loss $\mathcal{L}_{c}$ exploits a calibration receiver (available only during offline training) that provides the target waveform $\mathbf{y}^{(i)}$. Given the predicted delay $\widehat d^{(i)}$, a differentiable fractional-delay operator $\mathcal{S}_{\widehat d^{(i)}}$ aligns the attacker-side waveform $\mathbf{x}^{(i)}$:
\begin{equation}
\widetilde{\mathbf{x}}^{(i)}
=
\mathcal{S}_{\widehat d^{(i)}}
\left\{
\mathbf{x}^{(i)}
\right\}.
\end{equation}
The complex coefficient that minimizes the residual at the calibration receiver is
\begin{equation}
c_{\mathrm{opt}}^{(i)}
= - \frac{
\left(
\widetilde{\mathbf{x}}^{(i)}
\right)^{H}
\mathbf{y}^{(i)}
}{
\left(
\widetilde{\mathbf{x}}^{(i)}
\right)^{H}
\widetilde{\mathbf{x}}^{(i)}
},
\end{equation}
and the cancellation loss is defined as the normalized residual power after optimal cancellation:
\begin{equation}
\mathcal{L}_{c}
=
\frac{1}{N_{\mathrm{tr}}}
\sum_{i=1}^{N_{\mathrm{tr}}}
\log
\left[
1+
\frac{
\left\|
\mathbf{y}^{(i)}
+
c_{\mathrm{opt}}^{(i)}
\widetilde{\mathbf{x}}^{(i)}
\right\|_{2}^{2}
}{
\left\|
\mathbf{y}^{(i)}
\right\|_{2}^{2}
}
\right].
\end{equation}
The weighting $\lambda_c(e)$ is linearly annealed from $0.05$ to $3.0$ over the first $150$ epochs, allowing the network to first learn the supervised mapping and then progressively emphasize the cancellation objective. The calibration data are used exclusively during offline training and are not available during online inference, preserving the victim-blind operation.

\subsubsection{Delay-Derived Carrier-Phase Correction}

The neural networks predict only the temporal delay; the corresponding carrier-phase correction is derived deterministically. Given the final delay estimate $\widehat d$ in samples, the delay in seconds is $\widehat\tau = \widehat d / f_s$. The associated phase rotation is
\begin{equation}
\widehat\phi_c
=
\operatorname{wrap}_{[-\pi,\pi)}
\left(
\phi_0
-
2\pi f_c
\frac{\widehat d}{f_s}
\right),
\end{equation}
where $\phi_0$ is an initial calibrated RF-chain phase offset. This transformation converts the learned temporal alignment into the phase correction required for coherent RF cancellation. During online operation, the normalized I/Q window is passed through the two-stage network to obtain $\widehat d$; the waveform is then fractionally shifted by $\widehat d$ samples and rotated by $\widehat\phi_c$ before polarity inversion and transmission, with no reliance on calibration-receiver data.

\begin{figure}[t!]
    \centering
    \includegraphics[width=0.9\linewidth]{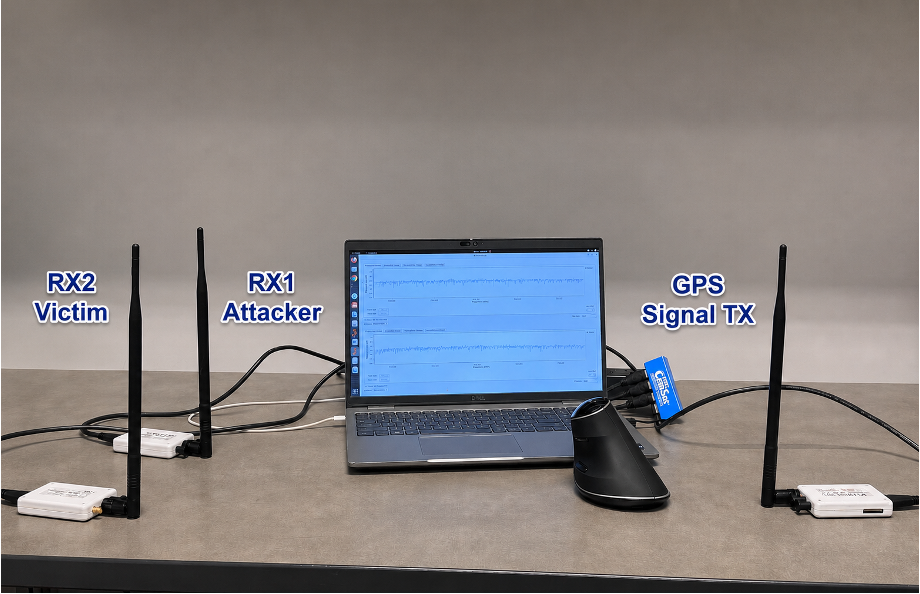}
    \caption{Experimental setup for the System model.}
    \label{fig:Exp_Setup}
\end{figure}

\begin{figure}[t!]
    \centering
    \includegraphics[width=0.9\linewidth]{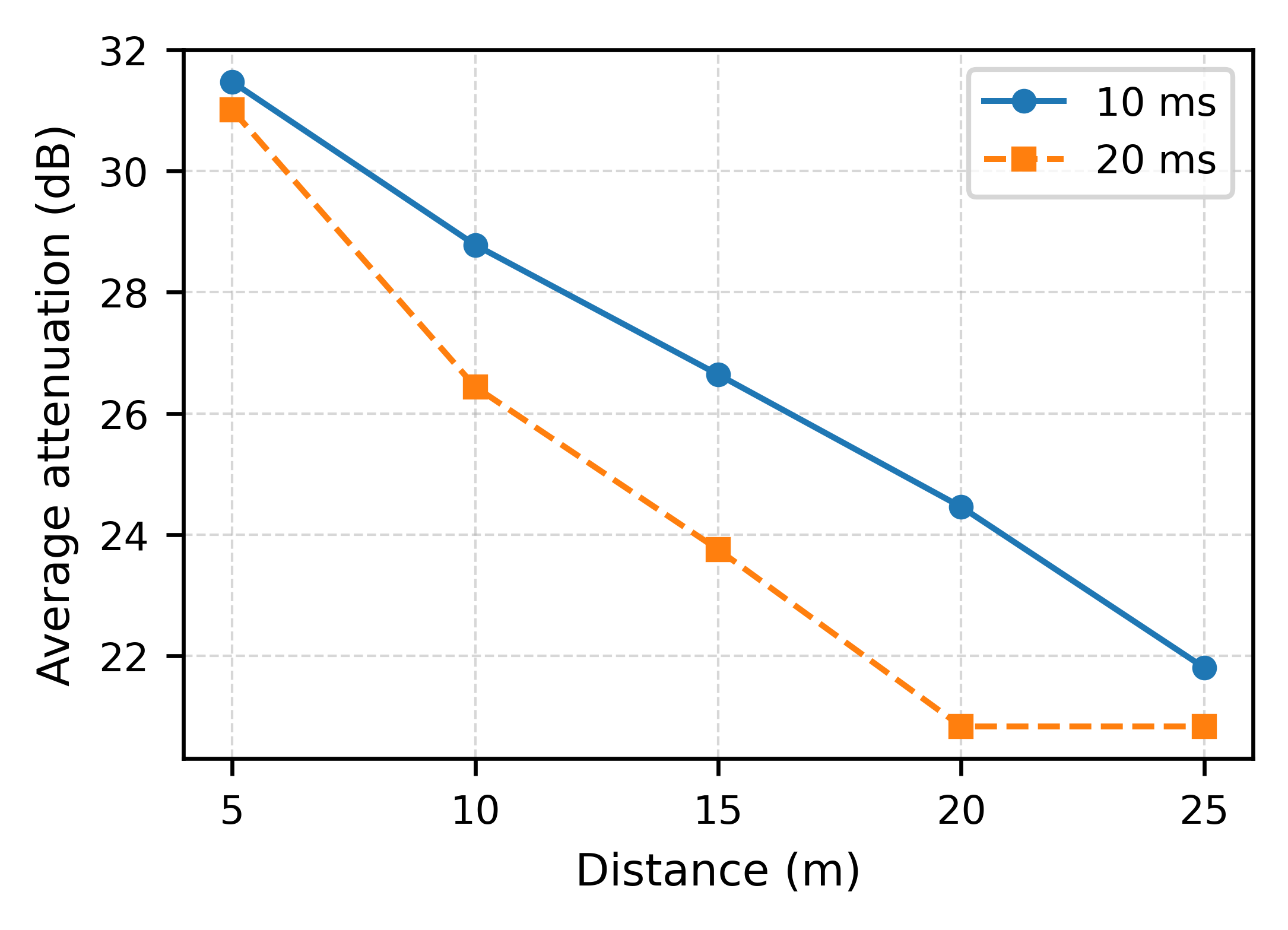}
    \caption{Achieved average attenuation over different distances.}
    \label{fig:attvsdist}
\end{figure}

\subsubsection{Network and Training Configuration}

The complete estimator follows a coarse-to-fine processing sequence. The normalized I/Q window is first passed through Stage~1 to obtain the full-delay estimate \(\widehat d_{1}\), which is rounded to produce the integer component \(\widehat m\). Stage~2 then uses the same normalized I/Q observations together with \(\widehat m\) to estimate the residual fractional delay \(\widehat\alpha\). The final delay is reconstructed as

\begin{equation}
\widehat d
=
\widehat m+\widehat\alpha,
\end{equation}

and is subsequently used for fractional waveform alignment and the deterministic carrier-phase correction. The calibration-receiver waveform contributes only to the offline cancellation-aware objective; online inference requires only the attacker-side I/Q window.

Table~\ref{tab:delay_learning_parameters} summarizes the principal data, network, and training parameters used in the implementation.

\begin{table}[t]
\centering
\caption{Two-Stage Delay-Estimation Parameters}
\label{tab:delay_learning_parameters}
\renewcommand{\arraystretch}{1.08}
\setlength{\tabcolsep}{4pt}
\begin{tabular}{ll}
\hline
\textbf{Parameter} & \textbf{Value} \\
\hline

Sampling rate & \(2.1~\mathrm{MHz}\) \\
Observation duration & \([10,20]~\mathrm{ms}\) \\
Data split & \(70\%/15\%/15\%\) \\
Split guard & [10,20] windows \\
Stage-1 stem & 64 channels, \(k=15\), \(s=2\) \\
Stage-1 residual blocks & 8, \(k=7\), dilations \(1,2,4,8\) \\
Stage-1 batch size & 64 \\
Stage-1 learning rate & \(3\times10^{-4}\) \\
Stage-2 stem & 48 channels, \(k=15\), \(s=4\) \\
Stage-2 fine dilations & \(1,2,4,8,16,32\) \\
Stage-2 batch size & 16 \\
Stage-2 learning rate & \(8\times10^{-5}\) \\
Delay-loss weights & \(\lambda_{\alpha}=60,\ \lambda_d=30\) \\
Cancellation weight & \(0.05\rightarrow3.0\) over 150 epochs \\
Optimizer & AdamW \\
Gradient clipping & 2.0 \\
\hline
\end{tabular}
\end{table}

\section{Experimental Evaluation}
This section describes the hardware and software platform used to evaluate the proposed GPS signal nullification framework. We first validate the proposed methodology under ideal simulation conditions and then assess its performance using over-the-air measurements collected with software-defined radios (SDRs).

\subsection{Implementation System}
All experiments were conducted on a workstation equipped with an Intel Core i7 processor and 16 GB RAM running Ubuntu. The signal processing pipeline was implemented in Python 3.10, while the SDR devices were controlled using UHD GNU Radio v3.10.1.1.

The experimental setup consists of three Ettus Research USRP B205mini devices configured according to the system model shown in Figure \ref{fig:system_model}. As illustrated in Figure \ref{fig:Exp_Setup}, one USRP acts as the GPS transmitter, while two USRPs serve as the attacker (RX1) and victim (RX2) receivers. All devices are connected to the host computer via a powered USB hub. The transmitter broadcasts GPS L1 signals at 1575.42 MHz with a sampling rate of 2.1 MS/s and 8-bit I/Q quantization. The received signals at RX1 and RX2 are recorded for offline processing, anti-signal generation, and nullification analysis.

\begin{figure}
    \centering
    \includegraphics[width=0.9\linewidth]{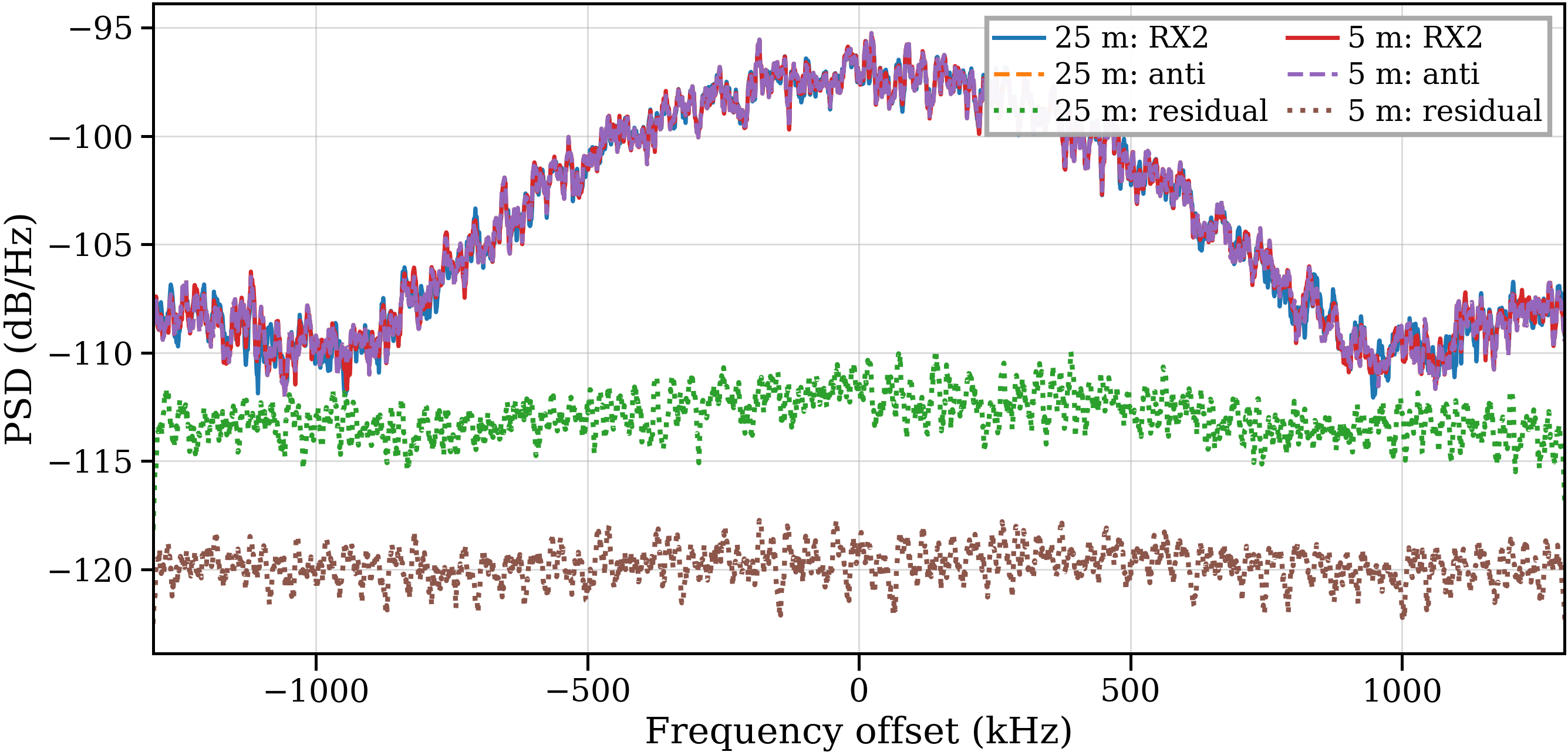}
    \caption{Power spectral density of victim, anti-signal, and residual for two distances under simulation setting.}
    \label{fig:sim_psd}
\end{figure}

\subsection{Experimental Analysis}
A baseline GPS dataset was generated using \texttt{gps-sdr-sim} together with publicly available RINEX ephemeris files to synthesize realistic GPS L1 baseband waveforms. The experimental evaluation consists of two stages. First, we validate the proposed nullification framework under ideal simulation conditions where channel impairments are absent. We then perform over-the-air experiments using the USRP platform to assess the practical effectiveness of the proposed approach under real wireless propagation.

\subsubsection{Simulation Experiment} \label{sec:simExp}
To establish the feasibility of the proposed nullification framework, we first consider an idealized simulation environment. Using \texttt{gps-sdr-sim}, two sets of 10-second GPS signal streams were generated corresponding to the attacker and victim locations separated by distances of $d\in\{5,10,15,20,25\}$~m. The attacker's received signal is used to construct the anti-signal through the proposed BAF framework.

Assuming an ideal propagation channel between the attacker and victim, the only parameter requiring estimation is the relative sample delay between the anti-signal and the victim signal. Once this delay is estimated, the anti-signal is time-shifted and combined with the victim signal to maximize destructive interference and minimize the residual signal power.

To estimate the delay, both signals are divided into non-overlapping windows of 10 and 20~ms. Each victim window is cross-correlated with its corresponding anti-signal window to determine the delay that maximizes alignment. The resulting delay is then used to generate the residual signal, and the corresponding attenuation is recorded. This procedure is repeated for all distances and both window sizes.

Figure \ref{fig:attvsdist} summarizes the average attenuation achieved under different attacker-victim separations. The proposed framework achieves an average of 31.4~dB attenuation at a separation of 5~m using 10~ms windows. As the separation increases, the attainable attenuation gradually decreases, with a noticeably steeper degradation for the 20~ms windows. This behavior suggests that shorter windows provide finer temporal resolution, enabling more accurate delay estimation and signals have to stay aligned for less time. For example, at 25~m the average attenuation decreases to approximately 22~dB for 10~ms windows and 20.7~dB for 20~ms windows.

To further illustrate the nullification effect, Figure \ref{fig:sim_psd} presents the power spectral density (PSD) of the victim signal, anti-signal, and residual signal for distances of 5~m and 25~m. In both cases, the residual spectrum exhibits substantial attenuation relative to the original signals, demonstrating the effectiveness of the proposed cancellation framework under ideal operating conditions.

Having established the feasibility of the proposed framework in simulation, we next evaluate its performance using real-world measurements collected with the USRP testbed.

\begin{figure}
    \centering
    \includegraphics[width=0.9\linewidth]{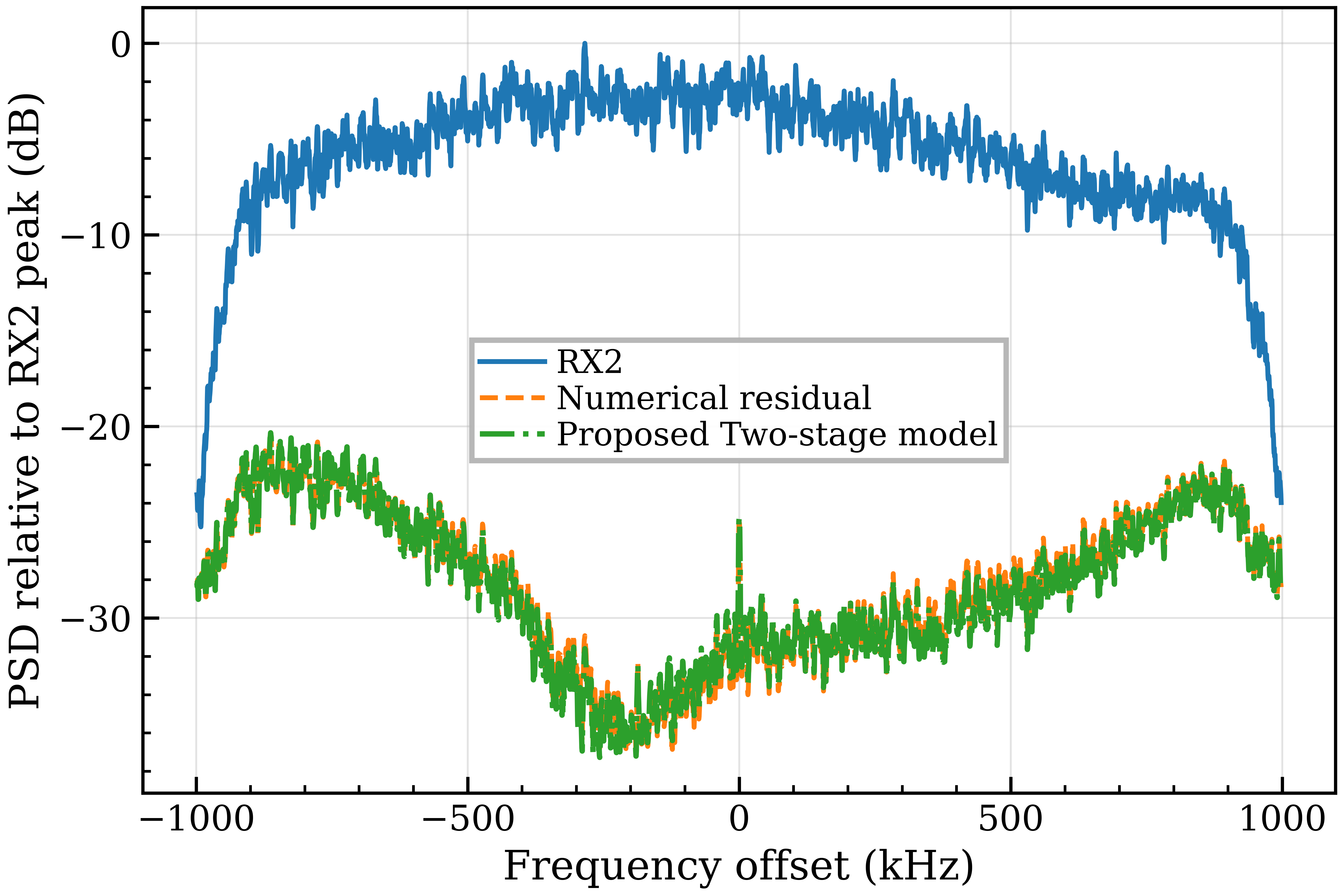}
    \caption{Power spectral density of victim, anti-signal, and residual under real data setting for 10 ms widow size.}
    \label{fig:real_psd_10}
\end{figure}

\begin{figure}
    \centering
    \includegraphics[width=0.9\linewidth]{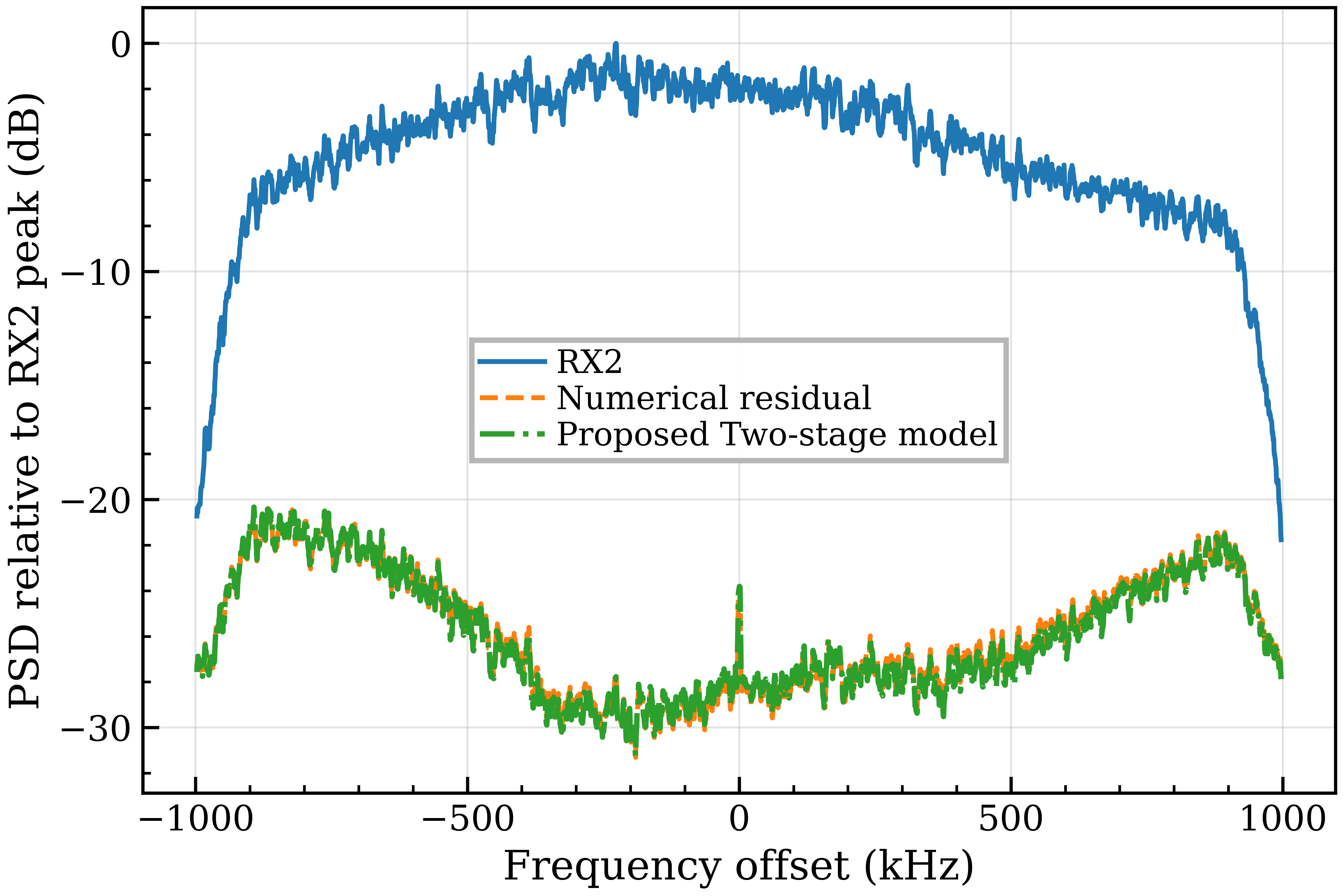}
    \caption{Power spectral density of victim, anti-signal, and residual under real data setting for 20 ms widow size.}
    \label{fig:real_psd_20}
\end{figure}

\subsubsection{Real Data Experiment}
The real-world evaluation was conducted using the USRP testbed shown in Figure \ref{fig:Exp_Setup}. A 10-second GPS signal corresponding to a fixed location on the University of Nebraska--Lincoln campus was transmitted using the GPS transmitter. The attacker (RX1) and victim (RX2) receivers were positioned approximately four feet apart while maintaining an equal distance from the transmitter, forming an equilateral triangle.

The received signals from RX1 and RX2 were recorded and processed following the same pipeline used in the simulation study. The attacker signal was first converted into an anti-signal, after which both the anti-signal and victim signal were partitioned into non-overlapping windows of 10 and 20~ms. Cross-correlation was then used to estimate the sample delay required to maximize destructive interference. Each pair of signal windows together with the corresponding optimal delay formed one training sample for the proposed deep learning model, whose objective is to predict the required delay directly from the victim and anti-signal waveforms.

The resulting dataset was partitioned chronologically into 70\%, 15\%, and 15\% training, validation, and testing subsets, respectively. The model was trained using the Adam optimizer with a learning rate of $10^{-3}$, a batch size of 32, and 150 training epochs. During inference, the predicted delay was applied to the anti-signal prior to combining it with the victim signal, producing the residual signal.

Figures \ref{fig:real_psd_10} and \ref{fig:real_psd_20} present the PSD of the victim signal, anti-signal, and the resulting residual signal for a representative test example at window size of 10 and 20 ms, respectively. Significant attenuation is observed across most of the 2.1~MHz GPS bandwidth, particularly near the center frequency. In contrast, attenuation near the band edges is comparatively weaker. This behavior is expected since practical wireless channels introduce frequency-dependent phase and amplitude distortions, preventing perfect destructive interference across all spectral components.

\begin{figure}[t]
    \centering
    \includegraphics[width=0.9\linewidth]{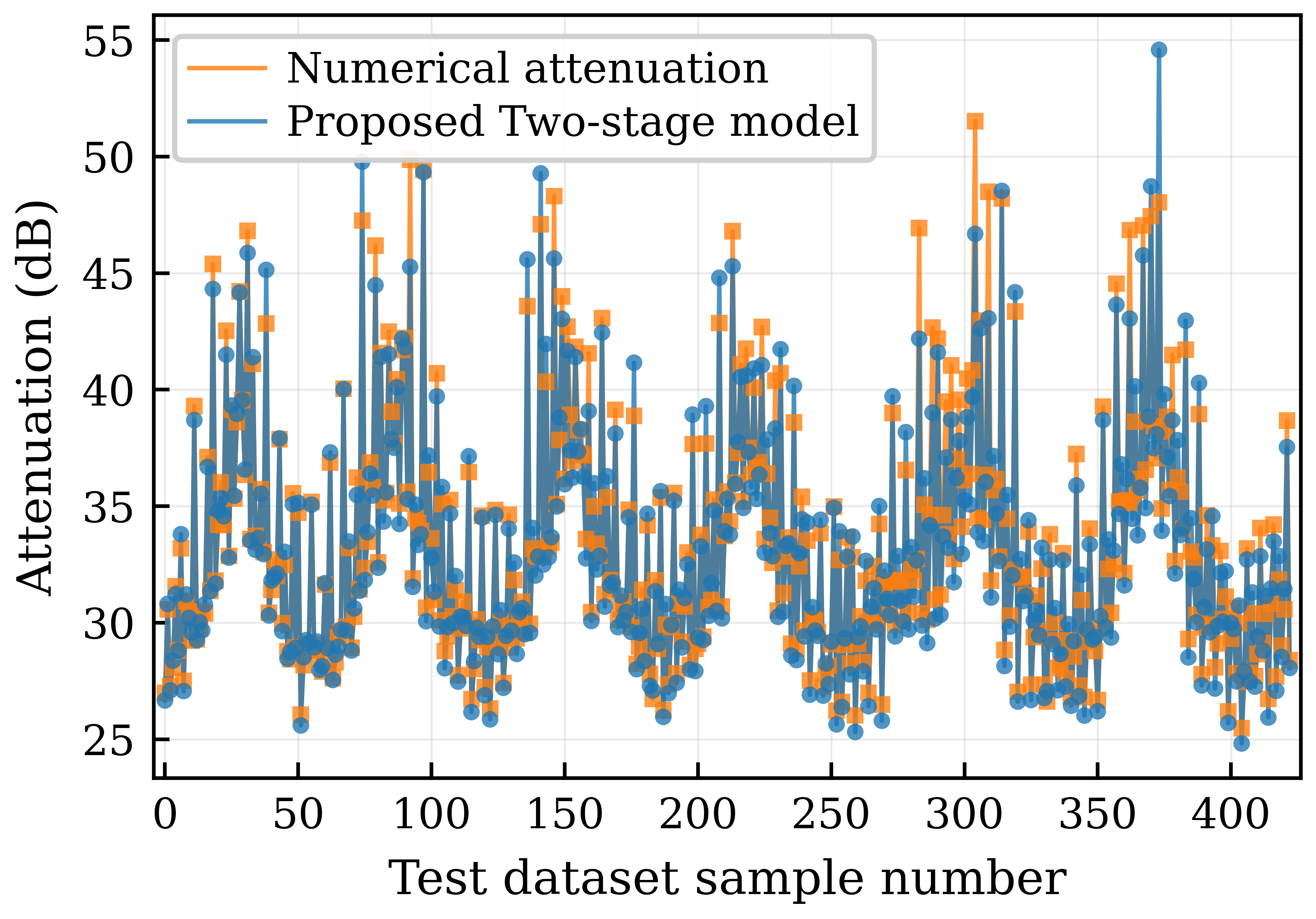}
    \caption{Comparison of the attenuation obtained using the numerical reference method and the proposed two-stage model for each test dataset sample.}
    \label{fig:real_sam_vs_atten_10}
\end{figure}

The proposed model achieved mean absolute errors (MAEs) of 0.003599 and 0.006184 samples and root-mean-square errors (RMSEs) of 0.003902 and 0.006365 samples for the 10~ms and 20~ms window configurations, respectively. Figure \ref{fig:real_sam_vs_atten_10} compares the attenuation obtained using the predicted delays with that obtained using the numerical reference delays across the test set. The close agreement between the two attenuation profiles indicates that the proposed two-stage model accurately estimates the optimal delay while maintaining substantial signal attenuation.

\section{Conclusion} \label{sec:conclusion}

This paper presented \textit{GhostLock}, a learning-assisted blind GPS signal nullification framework that suppresses authentic GPS signals without requiring victim-side observations or synchronization information. By combining a machine learning-based synchronization module with blind adaptive filtering, GhostLock reconstructs coherent GPS anti-signals using only the attacker's local observations and an approximate estimate of the victim's location. We implemented the proposed framework on an SDR-based platform and validated its performance through both simulation and over-the-air experiments, demonstrating its ability to accurately estimate synchronization parameters and effectively disrupt GPS signal acquisition under realistic attacker constraints. These results highlight a previously unexplored GPS threat model and underscore the need for future acquisition-stage detection and mitigation mechanisms capable of defending against blind GPS signal nullification attacks.


\bibliographystyle{IEEEtran}
\bibliography{bibfile}

\end{document}